\documentclass[a4paper,aps,prl,twocolumn,floats,showpacs]{revtex4}
\usepackage{graphicx}
\usepackage{amsmath,amssymb,amsfonts}
\usepackage[dvips]{hyperref}
\begin{document}

\title{Light asymmetric dark matter from new strong dynamics}
\author{Mads T. Frandsen}
\email{m.frandsen1@physics.ox.ac.uk}
\author{Subir Sarkar}
\author{Kai Schmidt-Hoberg}
\affiliation{Rudolf Peierls Centre for Theoretical Physics, University
  of Oxford, Oxford OX1 3NP, United Kingdom}

\begin{abstract} 
  A $\sim\!5$ GeV `dark baryon' with a cosmic asymmetry similar to
  that of baryons is a natural candidate for the dark matter. We study
  the possibility of generating such a state through dynamical
  electroweak symmetry breaking, and show that it can share the relic
  baryon asymmetry via sphaleron interactions, even though it has no
  electroweak interactions. The scattering cross-section on nucleons,
  estimated in analogy to QCD, is within reach of underground direct
  detection experiments.
\end{abstract}

\date{\today}
\pacs{95.35.+d,12.60.Nz} 
\maketitle

\section{Asymmetric dark matter}

A particle-antiparticle asymmetry in dark matter, similar to that in
baryons, would provide a natural link between their observed
abundances. A classic example of such \emph{asymmetric} dark matter
(ADM) is the lightest neutral technibaryon (TB)
\cite{Nussinov:1985xr,Chivukula:1989qb,Barr:1990ca} with a mass of
$\mathcal{O}(1)$ TeV in technicolour (TC) models of electroweak (EW)
symmetry breaking \cite{Weinberg:1979bn}. Other techni-interacting
massive particles (TIMPs) \cite{Gudnason:2006ug,Ryttov:2008xe} have
been considered, some of which are pseudo Nambu-Goldstone bosons
(pNGBs) of the TC interactions, hence lighter with mass of
$\mathcal{O}(100)$ GeV.

While gravitational instability in collisionless cold dark matter
(CDM) explains structure formation on large scales well, the predicted
substructure on galactic scales is at variance with observations
suggesting that CDM may be \emph{self-interacting}
\cite{Spergel:1999mh}. If ADM arises in a strongly coupled theory as a
\emph{composite} particle $\chi$ then it would naturally have
self-interactions, with a cross-section large enough to address the
small scale structure issue if their strong-interaction scale is of
${\cal O}$(GeV) (assuming the scaling resembles that in QCD). The
self-annihilation cross-section would naturally be of the same order,
ensuring that no significant symmetric abundance survives from the
early universe. The relic abundance is then given simply by
$\Omega_\chi = (m_\chi \mathcal{N}_\chi/m_\mathrm{B}
\mathcal{N}_\mathrm{B})\Omega_B$ where $\mathcal{N}_{\mathrm{B},
  \chi}$ are the respective asymmetries of baryons and ADM ({\it e.g.}
$\mathcal{N}_\mathrm{B} \equiv (n_\mathrm{B} -
n_\mathrm{\bar{B}})/(n_\mathrm{B} + n_\mathrm{\bar{B}})$). Now if some
process ensures $\mathcal{N}_\chi \sim \mathcal{N}_\mathrm{B}$ then
the observed cosmological dark matter abundance is realised for a
$\sim\!5$ GeV $\chi$ particle. Excitingly, recent signals in the
underground direct detection experiments DAMA \cite{Bernabei:2010mq}
and CoGeNT \cite{Aalseth:2010vx} have been interpreted in terms of
such light dark matter \cite{Kopp:2009qt} and generated renewed
interest in GeV scale ADM from new strong dynamics
\cite{Kaplan:1991ah}. While subsequent experiments like XENON100
\cite{Aprile:2010um} and CDMSII \cite{Akerib:2010pv} have not
confirmed these claims, their results still allow $\sim\!5$ GeV CDM
with a spin-independent scattering cross-section on nucleons as high
as $\sim\!10^{-39}$~cm$^2$. Moreover such particles can be accreted by
the Sun in large numbers (since they cannot annihilate) and affect
heat transport so as to measurably alter the fluxes of low energy
neutrinos \cite{Frandsen:2010yj}.

In this letter we consider a mechanism for generating light asymmetric
dark matter by extending the standard $SU(3)_\mathrm{c} \otimes
SU(2)_\mathrm{L} \otimes U(1)_Y$ model (SM) to include a new strong
interaction exhibiting two sectors $S_1$, $S_2$:\\
i) $S_1$ breaks EW symmetry dynamically at a scale $\Lambda_1$ and the
composite spectrum includes $\mathcal{O}(1)$ TeV mass particles TB carrying
a new global $U(1)_\mathrm{TB}$, just like `technibaryons' in
technicolour \cite{Weinberg:1979bn}. The constituents of TB carry weak
quantum numbers such that TB couples to the fermion number violating
EW sphaleron interactions which distribute any pre-existing fermion
asymmetry between baryons and TB's, ensuring that:
\begin{equation}
\label{darkpath1}
\mathcal{N}_\mathrm{TB}  \sim \mathcal{N}_\mathrm{B}.
\end{equation}
ii) $S_2$ is uncharged under the SM and becomes strongly interacting
at a scale $\Lambda_2$ of ${\cal O}$(GeV); its spectrum includes
(composite) few GeV mass particles $\chi$ also carrying the global
$U(1)_\mathrm{TB}$ quantum number. The two sectors are coupled via
operators that induce $U(1)_\mathrm{TB}$ preserving fast decays:
$\mathrm{TB} \to \chi + X$.  This interaction keeps $\chi$ in
equilibrium with T down to $T \lesssim0.1m_\mathrm{TB}$, {\it i.e.}
below the temperature $T_\mathrm{sph} \sim m_W$ at which the sphaleron
interactions `freeze-out'. Thus the sphaleron induced asymmetry in TB
is converted into a similar asymmetry in $\chi$:
\begin{equation}
\label{darkpath2}
\mathcal{N}_\chi \sim \mathcal{N}_\mathrm{TB}  \  .
\end{equation}
This naturally connects the relic density of light ADM to the relic
density of baryons via sphalerons, \emph{without} requiring $\chi$ (or
its constituents) to carry EW quantum numbers. Thus experimental
constraints on such light particles from the $W$ decay width {\it etc}
are not relevant.

In fact all that is necessary to generate light ADM via sphalerons is
EW charged states at the weak scale which can decay rapidly into ADM.
However we consider it more appealing to link ADM to dynamical EW
symmetry breaking and consider the possibility that the two sectors
above arise from a \emph{single} strongly interacting (technicolour)
extension of the SM which develops two different dynamical scales
$\Lambda_{1,2}$. 

This can happen {\it e.g.} if the theory contains fermions in
\emph{different} representations of the TC gauge group since the
critical value $\alpha_\mathrm{c}$ of the coupling which breaks chiral
symmetry depends on the quadratic Casimir $C_2$ of the representation
(using a simple one-gluon exchange estimate)
\cite{Raby:1979my,Marciano:1980zf}.  This possibility has been
considered in `two-scale' TC \cite{Lane:1989ej} --- a variant of the
earlier idea \cite{Marciano:1980zf} that if QCD contains fermions in a
higher-dimensional representation, then these might dynamically break
the EW symmetry at the correct scale.  A second possibility is if
large four-fermion operators are present as in the gauged
Nambu-Jona-Lasanio (NJL) model \cite{Nambu:1961tp}.  If the
four-fermion coupling is sufficiently large, this interaction can
drive chiral symmetry breaking, allowing a much smaller value of the
critical gauge coupling as in `top quark condensation'
\cite{Miransky:1988xi} or `topcolour' \cite{Hill:1994hp} models. If
only some of the fermions participate in the four-fermion
interactions, a scale separation can arise. SM singlet techni-fermions
have been introduced earlier \cite{Dietrich:2005jn} in models of
`Minimal Walking Technicolour' \cite{Sannino:2004qp} and `Conformal
Technicolor' \cite{Luty:2004ye}, in order to achieve (near-) conformal
or `walking' dynamics, while still maintaining a minimal sector which
breaks EW symmetry.

\section{Two scales from a strongly interacting theory}

We consider a non-Abelian gauge theory with gauge group $G$: $N_1$
fermions transforming according to a representation $\mathcal{R}_1$ of
$G$ are gauged under the EW symmetry, and $N_2$ fermions transforming
according to a representation $\mathcal{R}_2$ of $G$ are SM singlets.

Using the ladder approximation to the Schwinger-Dyson equations, the
critical value of the coupling for chiral symmetry breaking is
\cite{Appelquist:1988yc}:
\begin{equation}
\alpha_\mathrm{c} = \frac{\pi}{3C_2(\mathcal{R})}.
\end{equation}
We now take the gauge group to be $SU(N_\mathrm{C})$ and
consider representations such that $C_2(\mathcal{R}_1)\geq
C_2(\mathcal{R}_2)$. Integrating the one-loop beta-function 
$\beta(\alpha) = -(\frac{\alpha^2}{2\pi}\beta_0 +
\frac{\alpha^3}{8\pi^2}\beta_1 + \ldots)$ from
$\Lambda_1$ to $\Lambda_2$ yields the ratio of the scales:
\begin{equation}
  \frac{\Lambda_1}{\Lambda_2} \simeq 
  \exp\left[\frac{2\pi}{\beta_0(\mathcal{R}_2)} 
    \bigg(\alpha_\mathrm{c}(\mathcal{R}_1)^{-1} 
  - \alpha_\mathrm{c}(\mathcal{R}_2)^{-1} \bigg)
  \right] \ ,
\label{Eq:scaleratio}
\end{equation}
Since $\Lambda_1 \geq \Lambda_2$,or equivalently
$\alpha_\mathrm{c}(\mathcal{R}_1) \leq
\alpha_\mathrm{c}(\mathcal{R}_2)$, the fermions in the representation
$\mathcal{R}_1$ are, in this approximation, decoupled below
$\Lambda_1$, so only $\beta_0(\mathcal{R}_2)$ appears in the
exponent. If $\beta_0(\mathcal{R}_2)$ and $\alpha_\mathrm{c}$ are
small then the scale separation can be large, {\it i.e.} we can have
$\Lambda_1 \sim \Lambda_\mathrm{TC}$ and $\Lambda_2 \sim
\Lambda_\mathrm{ADM}$. Our estimate of $\alpha_\mathrm{c}$ should be
compared to the two-loop fixed point value of the coupling:
\begin{equation}
\alpha_*=-4\pi\frac{\beta_0}{\beta_1}.
\end{equation}
If $\alpha_*<\alpha_\mathrm{c}$ the theory will run to an infra-red
fixed point before triggering chiral symmetry breaking. The lower
boundary of the conformal window is thus identified by demanding
$\alpha_*(\mathcal{R}_1,\mathcal{R}_2)=\alpha_\mathrm{c}(\mathcal{R}_1)$.

For the fermions transforming under $\mathcal{R}_1$ to break the EW
symmetry at $\Lambda_1$ they must be charged under the EW gauge
group. The minimal choice (dictated also by constraints from EW
precision measurements \cite{Ryttov:2008xe}) is $N_1=2$ Dirac flavors
with the left-handed Weyl spinors arranged in an
$SU(2)_\mathrm{L}\otimes U(1)_Y$ weak doublet $Q_\mathrm{L}$, along
with $N_2$ SM singlet Dirac flavors $\lambda$:
\begin{eqnarray} 
  Q_\mathrm{L}^a &=& \left(\begin{array}{c} U^{a} \\
      D^{a} \end{array}\right)_\mathrm{L},\ 
  U_\mathrm{R}^a,\ D_\mathrm{R}^a,\  \lambda^b \nonumber \\
  a &=& 1, \ldots d(\mathcal{R}_1),\ b=1, \ldots d(\mathcal{R}_2).
\end{eqnarray} 
The condensate $\langle U_\mathrm{L} U^*_\mathrm{R} + D_\mathrm{L}
D^*_\mathrm{R} + \mathrm{h.c.} \rangle$ breaks EW symmetry and the TBs
are made out of the $U$s and $D$s while the dark matter candidate
$\chi$ is made of the $\lambda$s (and is a fermion or boson depending
on $N_\mathrm{C}$). The unbroken global symmetries of the TC
interactions (depending on $\mathcal{R}_{1,2}$) keep $\chi$ stable and
are discussed further below.

\begin{figure}[htp!]
{\includegraphics[width=0.67\columnwidth]{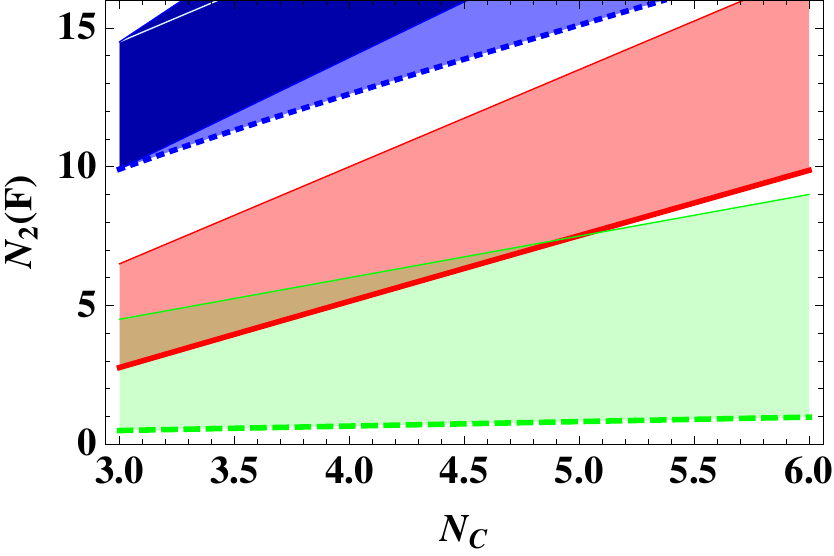}
\includegraphics[width=0.67\columnwidth]{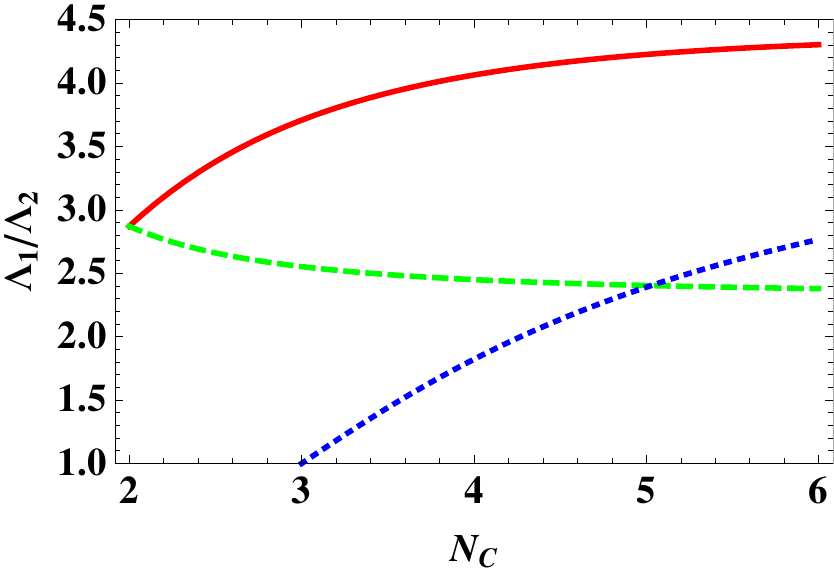}
}
\caption{Upper panel: The conformal window for $SU(N_\mathrm{C})$
  gauge theories with $N_2(F)$ fermions in the fundamental
  representation ($\mathcal{R}_2=F$), and 2 flavours in the (top to
  bottom): fundamental (dark blue), two-index antisymmetric (light
  blue), two-index symmetric (red), and adjoint (green)
  representations. Lower panel: The corresponding scale separation
  using the value of $N_2(F)$ at the lower boundary of the
  corresponding conformal window for the: two-index antisymmetric
  (blue dotted), adjoint (green dashed), and two-index symmetric (red
  solid) representations.}
\label{fig1}
\end{figure}

In Fig.~\ref{fig1} we show the conformal phase-diagram in the $(N_2,
N_\mathrm{C})$ plane, as well as the corresponding scale separation,
with $N_2$ taking the value along the lower boundary of the conformal
window. It is seen that we \emph{cannot} achieve large scale
separations without having to increase $N_2$ (\emph{i.e.}\! reduce
$\beta_0 (\mathcal{R}_2)$) to values where the theory is actually
IR-conformal. Below $\Lambda_1$, where the $\mathcal{R}_1$ fermions
decouple, the coupling simply runs too fast.

We consider now the effects of four-fermion operators, expected in any
case in a more complete theory {\it e.g.} when embedding the
technicolour gauge group in an extended technicolour (ETC) gauge group
\cite{Dimopoulos:1979es}, which is required to communicate EW breaking
to the SM fermions without introducing fundamental scalars. It is
assumed to break at some high scale $\Lambda_\mathrm{ETC}$, so in
integrating out the heavy ETC gauge bosons, four-fermion vector
currents appear at the TC scale.  This can also alleviate the tension
with the $S$-parameter \cite{Chivukula:2005bn}. Four-fermion operators
can also be induced by the non-perturbative dynamics of walking
technicolour itself \cite{Kurachi:2007at}.

\begin{figure}[htp!]
{
\includegraphics[width=0.67\columnwidth]{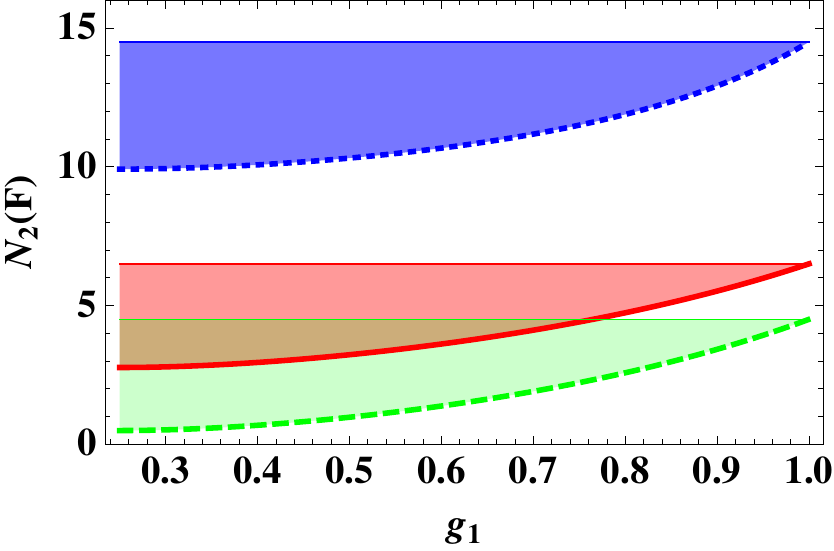}
\includegraphics[width=0.67\columnwidth]{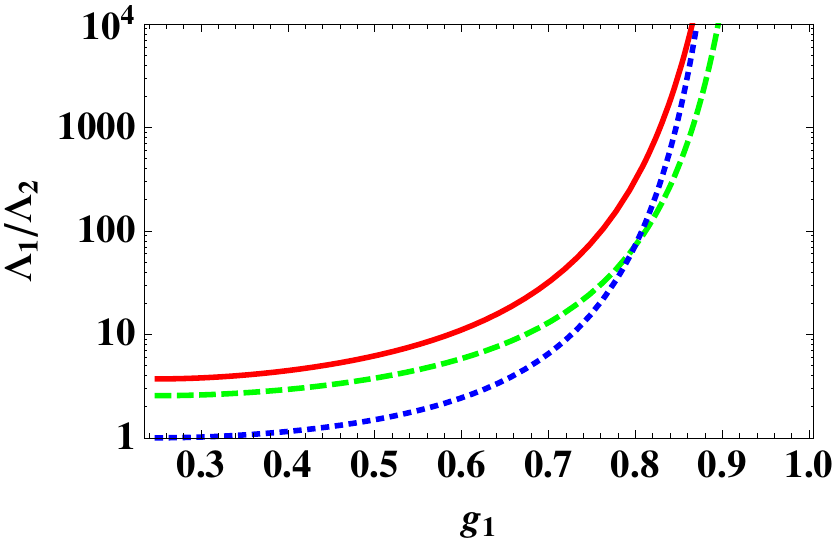}
}
\caption{As in Fig.~\ref{fig1}, but now as a function of $g_1$ (\ref{NJLcomeon}). Here
  the fundamental and antisymmetric representations are identical.}
\label{fig2}
\end{figure}

We adopt the gauged NJL model \cite{Nambu:1961tp} as a representative
simple theory with four-fermion interactions which leads to chiral
symmetry breaking.  While the four-fermion operator coupling is of the
form $VV-AA$ here (where $V$ and $A$ denote vector and axial vector
currents), the relation between chiral symmetry breaking on the one
hand and the gauge and four-fermion coupling on the other hand is
qualitatively the same in other cases as well. The gauged NJL model
for the fermions $Q$ transforming under $\mathcal{R}_1$ is defined by
\begin{eqnarray}
  \label{NJLcomeon}
  {\cal L} &=& \bar{Q} i {\not}{D} Q - 
  \frac{1}{4} \mathrm{Tr} F^a_{\mu \nu}F^{a\mu\nu}\,  \nonumber \\ 
  &+&\frac{4\pi^2g_1}{\Lambda^2 N_1 d(\mathcal{R}_1)} \left[ (\bar{Q} Q)^2 
    + (\bar{Q} i \gamma_5 T^a Q)^2 \right],  
    \end{eqnarray}
with $g_1$ the dimensionless four-fermion coupling and $\Lambda$ the
effective cut-off of the the NJL model. The critical line in the
$(\alpha, g_1)$-plane is now given by
\cite{Kondo:1988qd,Appelquist:1988fm}:
\begin{equation}
\alpha_\mathrm{c} (\mathcal{R}_1, g_1)= 
\begin{cases}
4 \left( \sqrt{g_1} - g_1 \right) \times \alpha_\mathrm{c}(\mathcal{R}_1)  
& \text{for } \, \tfrac{1}{4} < g_1 \leq 1\,, \\ 
\alpha_\mathrm{c}(\mathcal{R}_1) & \text{for } \, 0 \leq g_1 <\tfrac{1}{4}\,.
\end{cases} 
\nonumber
\end{equation}
The critical gauge coupling is thus reduced and in the limit $g_1 \to
1$, chiral symmetry breaking is driven purely by the four-fermion
interaction. The lower boundary of the conformal window, found by
imposing
$\alpha_*({\mathcal{R}_1,\mathcal{R}_2})=\alpha_\mathrm{c}(\mathcal{R}_1,
g_1)$, now changes \cite{Fukano:2010yv} as shown in Fig.~\ref{fig2} for
the case of $SU(3)$ gauge theories with $N_1=2$ and $\mathcal{R}_1$ in
different representations.  It is seen how the conformal window
shrinks as $g_1 \to 1$, allowing the addition of more matter
transforming under $\mathcal{R}_2$ while still having a theory that
breaks chiral symmetry. Fig.~\ref{fig2} also shows that the scale
separation $\Lambda_2/\Lambda_1$ can now be substantial.

Although we have used rather simple approximations, this shows
qualitatively how a significant scale separation can be achieved
within a single strongly interacting theory. Other dynamical
frameworks may allow large scale separations {\it e.g.} models with
fundamental scalars which condense at $\Lambda_1$ and thereby have a
similar effect to four-fermion operators.  Another possibility is
chiral gauge theories where some fermions and some gauge bosons
condense at $\Lambda_1$ resulting in a slowly evolving coupling below
$\Lambda_1$ and a large scale separation.

\section{Phenomenology}

The phenomenology will depend on the specific TC gauge group
and the fermion representations breaking EW symmetry,
but there are some generic implications.

If $\mathcal{R}_1=\mathcal{R}_2$ (both complex), there is a single
global $U(1)_{\rm TB}$ keeping $\chi$ stable.  We then expect to have
a sizeable Yukawa-type interaction between the heavy technibaryon TB,
the dark matter particle $\chi$ and {\it e.g}. the composite Higgs
which would keep the TB and $\chi$ particles in thermal equilibrium in
the early universe. However if $\mathcal{R}_1\neq \mathcal{R}_2$,
additional global $U(1)$'s in the theory can prevent such operators
from arising; in that case we would require additional interactions
such as ETC breaking these $U(1)$'s such that only the technibaryon in
the low scale sector remains stable \cite{Chivukula:1989qb}; an
interesting example of interactions allowing this has been given
\cite{Wilczek:1976qi}. We discuss explicit models elsewhere but
comment here on the implications for dark matter detection
experiments.

The symmetric component of $\chi$ can effectively annihilate into
states which do not carry the $U(1)_\mathrm{TB}$ ({\it e.g.}
technipions) and can subsequently decay to SM states. The asymmetric
component of $\chi$ cannot annihilate so there are no indirect
signatures (since decays are highly suppressed compared to TeV scale
ADM \cite{Nardi:2008ix}).

We expect isospin-0 scalar and vector mesons like a `techni-sigma' (or
composite Higgs) and a `techni-omega' in both sectors.  The
`techni-omega' will mix with the SM hypercharge field after EW
symmetry breaking and induce couplings of $\chi$ to the SM sector,
just as we expect the `techni-sigma' to couple to SM fermions via
effective Yukawa couplings (induced {\it e.g.} by ETC). This can lead
to exciting signals at the LHC \cite{Chivukula:1989rn}.

The spin-idependent elastic scattering cross-section on nucleons from
either scalar or vector boson exchange is
\begin{equation}
\sigma \sim g_{\chi}^2 g_{q}^2 \frac{\mu^2}{m^4} ,
\end{equation}
which is $\sim\!(10^{-32}-10^{-30}) g_q^2$ cm$^2$ for a mediator mass
of $m \sim 5-15$ GeV, a reduced mass $\mu \sim 1$ GeV and a coupling
$g_\chi \sim 1$ between $\chi$ and the mediator. The coupling $g_q$
arises from mixing between the light and heavy states, thus
parametrically $g_q \sim \Lambda_2/\Lambda_1$, and in addition it is
proportional to small couplings --- the $U(1)_Y$ coupling $g'$ and
fermion hypercharges in the `techni-omega' case, and the light SM
fermion Yukawa couplings in the composite Higgs case. Hence we expect
$g_q \lesssim 10^{-4}$, so the cross-section is within reach of direct
detection experiments which are sensitive to nuclear recoil energies
of $\mathcal{O}$(keV) characteristic of $\sim\!5$ GeV ADM.
  
While pNGB's themselves may carry $U(1)_\mathrm{TB}$
\cite{Gudnason:2006ug} and provide another way of generating light ADM
in simple models of dynamical EW symmetry breaking, their
self-interactions are derivatively suppressed at low energies. By
contrast, the `dark baryon' ADM state $\chi$ considered here is
expected to have large self-interactions and thus interesting
astrophysical signatures \cite{Spergel:1999mh,Frandsen:2010yj}.

\section*{Acknowledgements}

MTF acknowledges a VKR Foundation Fellowship. We thank John
March-Russell and, especially, Francesco Sannino for stimulating
discussions, and the European Research and Training Network
``Unification in the LHC era'' (PITN-GA-2009-237920) for partial
support.

\end{document}